\documentclass[twocolumn,showpacs,showkeys,preprintnumbers,amsmath,amssymb]{revtex4}
\usepackage{graphicx}
\usepackage{dcolumn}
\usepackage{bm}
\def\1{\mathchoice{\rm 1\mskip-4.2mu l}{\rm 1\mskip-4.2mu l}{\rm
        1\mskip-4.6mu l}{\rm 1\mskip-5.2mu l}}

\begin{document}
\title{Noncyclic Pancharatnam phase for mixed state SU(2) evolution \\in neutron
    polarimetry}
\author{J. Klepp, S. Sponar, Y. Hasegawa, E. Jericha and G.
Badurek}
 \affiliation{TU Vienna, Atominstitut, Stadionallee 2, 1020 Vienna,
            Austria}
\date{\today}
\begin{abstract}
We have measured the Pancharatnam relative phase for
    spin-1/2 states.
    In a neutron polarimetry experiment the
    minima and maxima of intensity modulations, giving the
    Pancharatnam phase, were determined. We also considered
    general SU(2) evolution for mixed states.
    The results are in good
    agreement with theory.
\end{abstract}
 \maketitle
\section{\label{sec:intro}Introduction}
In recent years much attention has been paid to the concept of
geometric phase. Since its discovery by Berry \cite{Berry1984} the
subject was widely expanded and subdued to several generalizations
\cite{ShapereWilczek1989}, e.g. non adiabatic
\cite{AharanovAnandan1987} and noncyclic \cite{SamuelBhandari1988}
evolutions as well as the off-diagonal case
\cite{ManiniPistolesi2000}. Ever since, a great variety of
experimental confirmations of the Berry phase and its peculiar
properties have been accomplished (see e.g.
\cite{TomitaChao1986,WeinfurterBadurek1990,BadurekEtAl1993,AllmanEtAl1997,HasegawaEtAl2001,HasegawaEtAl2003}).
In 1956 Pancharatnam defined
$\phi=\mbox{arg}\langle\psi_0|\psi\rangle$ as the phase acquired
during an entirely arbitrary evolution of a wave-function
\cite{Pancharatnam1956}. Provided the evolution takes place under
condition of parallel transport, $\phi$ can be identified with the
noncyclic geometric phase. Quite recently, also a concept of mixed
state phase was developed by Sj\"{o}qvist \emph{et al.}
\cite{SjoeqvistEtAl2000} due to its importance in quantum
computation. The theoretical predictions have been tested by Du
\emph{et al.} \cite{DuEtAl2003} and Ericsson \emph{et al.}
\cite{EricssonEtAl2004} using NMR and single-photon
interferometry, respectively. In this letter, we report a neutron
polarimetry experiment for measuring the Pancharatnam relative
phase of a spinor subdued to an arbitrary SU(2) evolution,
implementing the method described by Wagh and Rakhecha in
\cite{WaghRakhecha1995}. We test its extension to the mixed state
case put forward by Larsson and Sj\"{o}qvist
\cite{LarssonSjoeqvist2003}.
\section{\label{sec:purestates}The pure state case}
\begin{figure}
    \begin{center}
    \scalebox{0.7}{\includegraphics{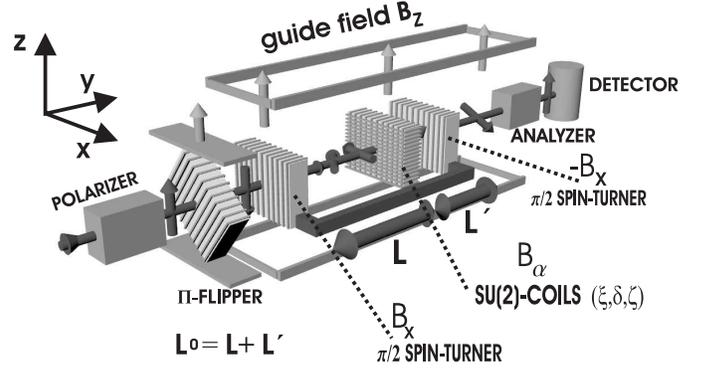}}
    \end{center}
    \caption{Sketch of the neutron polarimetry setup for
    Pancharatnam phase measurement
    including evolution of polarization vector for
    one specific relative position of
    spin turners at distance L$_0$ and SU(2) coils.}
    \label{fig1.eps}
\end{figure}
We consider the experimental setup shown in Fig. \ref{fig1.eps}. A
monochromatized neutron beam, polarized along the $+z$ direction
experiences a $\pi$/2 flip about the $x$-axis, therefore being
denoted as a superposition of the orthogonal states $|\pm
z\rangle$. The beam travels in an overall guide field B$_z$ of the
length L, which is chosen to be parallel to the $+z$ direction.
Under the influence of B$_z$ the polarization vector rotates
through an angle $\eta$ determined by the distance L. The two
orthogonal components $|\pm z\rangle$ of the rotated superposition
state acquire opposite phase undergoing the SU(2) transformation
$\hat {\mbox{U}}_0$, denoted by the matrix
\begin{eqnarray}\label{eq:SU(2)Matrix}
    \mbox{U}_0(\xi,\delta,\zeta)&=&
    \left(
        \begin{array}{cc}
            \mbox{e}^{i\delta}\cos\xi &
            ~-\mbox{e}^{-i\zeta}\sin\xi\\
            \mbox{e}^{i\zeta}\sin\xi & ~\mbox{e}^{-i\delta}\cos\xi
        \end{array}
    \right).
\end{eqnarray}
In neutron polarimetry the unitary, unimodular operator $\hat
{\mbox{U}}(\vec\alpha)$ is used for description of spin rotations
(transformations) carried out within magnetic fields. By
comparison of Eq.(\ref{eq:SU(2)Matrix}) to the matrix
representation of $\hat {\mbox{U}}(\vec\alpha)$, with
\begin{eqnarray}\label{eq:SpinRotationOperator}
    \hat {\mbox{U}}(\vec\alpha)&=&
    \exp\left(-i\frac{\hat{\vec\sigma}\cdot\vec\alpha}{2}\right)\nonumber\\
    &=&\hat{\1}\cos\frac{\alpha}{2}-i
    \left(\hat{\vec\sigma}\cdot\frac{\vec\alpha}{\alpha}
    \right)\sin\frac{\alpha}{2},
\end{eqnarray}
one is lead to sets ($\xi,\delta,\zeta$) of SU(2) parameters.
$\vec\alpha$ points to the direction of the magnetic field
constituting the rotation axis and $|\vec\alpha|=\alpha$ denotes
the rotation angle. $\hat{\vec\sigma}$ is the Pauli vector
operator. The resulting beam passes a distance L' along the guide
field B$_z$, its polarization rotating through the associated
angle $2\mbox{n}\pi - \eta$ (n$=1,2,...$). L$_0=$ L$+ $L' is set
in such a way that the accumulated rotation angle within B$_z$
equals an integer multiple of $2\pi$,
$\mbox{L}_0=\mbox{n}\pi\hbar\mbox{v}/|\mu \mbox{B}_z|$. Finally, a
$-\pi$/2 flip around the $x$-axis is applied and the output
intensity, calculated as
\begin{eqnarray}\label{eq:PureStateInt}
    \mbox{I}&=&
    \cos^2\xi\cos^2\delta+\sin^2\xi\sin^2(\zeta+\eta)
\end{eqnarray}
for the pure state case, is measured. The phase shift $\eta$,
implemented by variation of the distances L and L' at constant
L$_0$ causes intensity modulations, yielding the extreme values
\begin{eqnarray}
    \mbox{I}_{\mbox{\scriptsize min}}&=&
    \cos^2\xi\cos^2\delta,\\
    \mbox{I}_{\mbox{\scriptsize max}}&=&
    \cos^2\xi\cos^2\delta+\sin^2\xi.
\end{eqnarray}
In our case, the Pancharatnam phase $\phi$ is given by
substituting $|\psi_0\rangle$ by $|+z\rangle$ and $|\psi\rangle$
by $\hat {\mbox{U}}_0|+z\rangle$. The relative phase shift between
$|+z\rangle$ and $\hat {\mbox{U}}_0|+z\rangle$,
$\phi=\mbox{arg}\langle+z|\hat {\mbox{U}}_0|+z\rangle=
\delta+\mbox{arg}\cos\xi$  for the pure state case, is therefore
computed from
\begin{eqnarray}\label{eq:PureStatePhase}
    \phi&=&\arccos\sqrt\frac{\mbox{I}_{\mbox{\scriptsize min}}}
    {1-\mbox{I}_{\mbox{\scriptsize max}}+\mbox{I}_{\mbox{\scriptsize min}}}.
\end{eqnarray}
\section{\label{sec:mixedstatecase}The mixed state case}
More general, a neutron beam passing the setup explained above at
any arbitrary degree of polarization \emph{r} along the positive
$z$-axis is described by the density operator
\begin{eqnarray}\label{eq:densityoperator}
    \hat{\rho}&=&\frac{1}{2}(\hat{\1}+r\hat\sigma_z),
\end{eqnarray}
with the Pauli spin operator $\hat\sigma_z$ and $0\le r\le 1$. The
measured intensity for this mixed state is given by
\begin{eqnarray}\label{eq:MixedStateInt}
    \mbox{I}^{\rho}&=&\frac{1-r}{2}+r \mbox{I},
\end{eqnarray}
which obviously reduces to Eq.(\ref{eq:PureStateInt}) for $r=1$.
The relative mixed state phase $\Phi$ \cite{SjoeqvistEtAl2000} is
\begin{eqnarray}\label{eq:MixedStatePhase}
    \Phi&=&\arctan[r\tan(\delta+\arg\cos\xi)].
\end{eqnarray}
With the associated extreme values of Eq.(\ref{eq:MixedStateInt})
for $r\ge0$,
\begin{eqnarray}
    \mbox{I}^{\rho}_{\mbox{\scriptsize min}}&=&\frac{1-r}{2}+r\cos^2\xi\cos^2\delta\\
    \mbox{I}^{\rho}_{\mbox{\scriptsize max}}&=&\frac{1-r}{2}+
    r(\cos^2\xi\cos^2\delta+\sin^2\xi),
\end{eqnarray}
one obtains the following expression for the mixed state relative
phase \cite{LarssonSjoeqvist2003}:
\begin{widetext}
\begin{eqnarray}\label{eq:MixedStatePhInTermsOfImaxImin}
    \Phi&=&\arccos\sqrt{\frac{\mbox{I}^{\rho}_{\mbox{\scriptsize min}}-1/2(1-r)}
    {r^2[1/2(1+r)-\mbox{I}^{\rho}_{\mbox{\scriptsize max}}]+
    \mbox{I}^{\rho}_{\mbox{\scriptsize min}}-1/2(1-r)}}.
\end{eqnarray}
\end{widetext}
This formula is consistent with the pure state case since for
$r=1$ it reduces to Eq.(\ref{eq:PureStatePhase}).
The density
operator formalism is subject to an ambiguity, in a sense that
certain inherently distinct physical situations are described by
identical density matrices. This is often referred to as
'decomposition freedom' (see e.g. \cite{KultSjoeqvist2004}).
Moreover, it is not apparent from Eq.(\ref{eq:densityoperator}) by
what means some mixed state of the system should be produced.
Considering this, we introduced a $\pi$ flip in front of the
actual setup, resulting in a beam polarized in the $-z$ direction.
The intensities I$_{\mbox{\scriptsize off}}$ and
I$_{\mbox{\scriptsize on}}$, corresponding to 'spin flipper off'
and 'spin flipper on', respectively, were measured for equal time
intervals. A density matrix was calculated from a weighted sum of
I$_{\mbox{\scriptsize off}}$ and I$_{\mbox{\scriptsize on}}$
referring to a certain degree of polarization $r\ge 0$ and is
therefore equivalent to the associated mixed state. Since a real
polarizer does not work perfectly, it is important to note that
the incident neutron beam can already be considered to be in a
mixed state, as it was done in this work. For many experiments, in
contrast, it is sufficient to assume the incident beam to be
purely polarized in some arbitrarily chosen direction.
\section{\label{sec:experiment}Experiment}
\begin{figure*}
    \begin{center}
    \scalebox{0.6}{\includegraphics{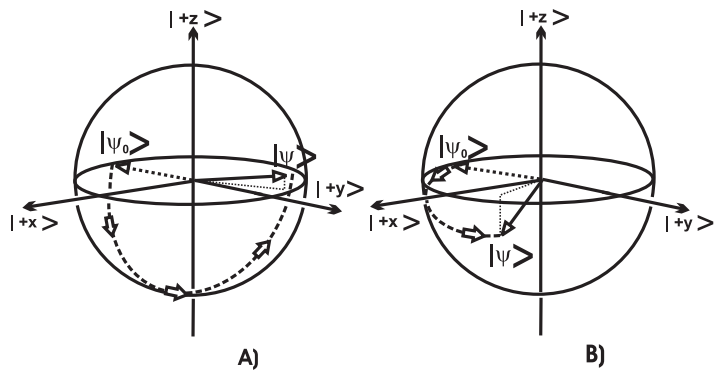}}
    \end{center}
    \caption{Rotations on the Poincar\'e sphere associated
    to parameter sets
    A: $(\xi^{\mbox{\scriptsize (A)}},\delta^{\mbox{\scriptsize (A)}},
    \zeta^{\mbox{\scriptsize (A)}})\rightarrow(1.71,0.38,-1.46)$ rad and B: $(\xi^{\mbox{\scriptsize (B)}},\delta^{\mbox{\scriptsize (B)}},
    \zeta^{\mbox{\scriptsize (B)}})\rightarrow(1.06,0.17,-1.40)$ rad
    for one specific relative position of spin turners and SU(2) coils.}
    \label{Fig2.eps}
\end{figure*}
The experiment was carried out at the tangential beam port of the
250 kW TRIGA research reactor of the Atomic Institute of the
Austrian Universities, Vienna. The neutron beam of mean wavelength
$\lambda=1.99$ \AA~, incident from a pyrolytic graphite
monochromator, was polarized along the $+z$ direction by
reflection from a bent Co-Ti supermirror array. All spin rotations
were implemented by Larmor precessions about the magnetic field
axes of DC coils made of anodized aluminium wire wound on frames
with rectangular profile. After passing the DC $\pi$ flipper the
beam encounters the first $\pi/2$ spin turner creating a
superposition of the orthogonal states $|\pm z\rangle$. The
polarization vector precesses in the $xy$-plane through the angle
$\eta$ around the $+z$ direction in the guide field B$_z$ and
length L at an angular frequency
$\omega=\frac{2\mu}{\hbar}\mbox{B}_z$. The guide field of length
L$_0$ was realized by two rectangular coils of 150 cm length along
the beam trajectory and a width of 12 cm. They were arranged in
Helmholtz geometry to provide for optimal homogeneity. The
combination of the guide field B$_z$ measured by a Hall probe
($=5.893\pm 0.022$ G) and the distance 4 L$_0$ (about 46 cm) was
chosen in such a way that more than one period ($\eta> 2\pi$) of
intensity oscillations could be scanned for its extreme values.
The subsequent SU(2) transformation $\mbox{U}_0(\xi,\delta,\zeta)$
was carried out by two coils with mutually orthogonal orientation
wound on the same frame. They define an arbitrary magnetic field
axis in the $xz$-plane, whose effect on the spin state is denoted
by Eq.(\ref{eq:SpinRotationOperator}). Along the distance L' from
the SU(2) coils to the second spin turner the polarization vector
precesses by an angle $2\mbox{n}\pi-\eta$ about the guide field
B$_z$ before being turned through $-\pi/2$ about the $x$-axis,
analyzed in the $+z$ direction by a second supermirror array and
finally counted by a $^3$He detector. The variation of the phase
shift $\eta$ is implemented by physical translation of the two
$\pi/2$ spin turners at constant distance L$_0=$ L $+$ L'. This
can be done conveniently by steps of some mm, exhibiting the
intensity oscillations evoked by the phase shift $\eta$.

Here, we present the results for two settings of SU(2) coil
currents. For each set of transformation parameters
($\xi,\delta,\zeta$), three minima and maxima of
$\mbox{I}_{\mbox{\scriptsize off}}$ and
$\mbox{I}_{\mbox{\scriptsize on}}$ were measured in one scan.
Given uncertainties contain both systematic and statistical
errors.
\begin{figure}
    \begin{center}
    \scalebox{0.7}{\includegraphics{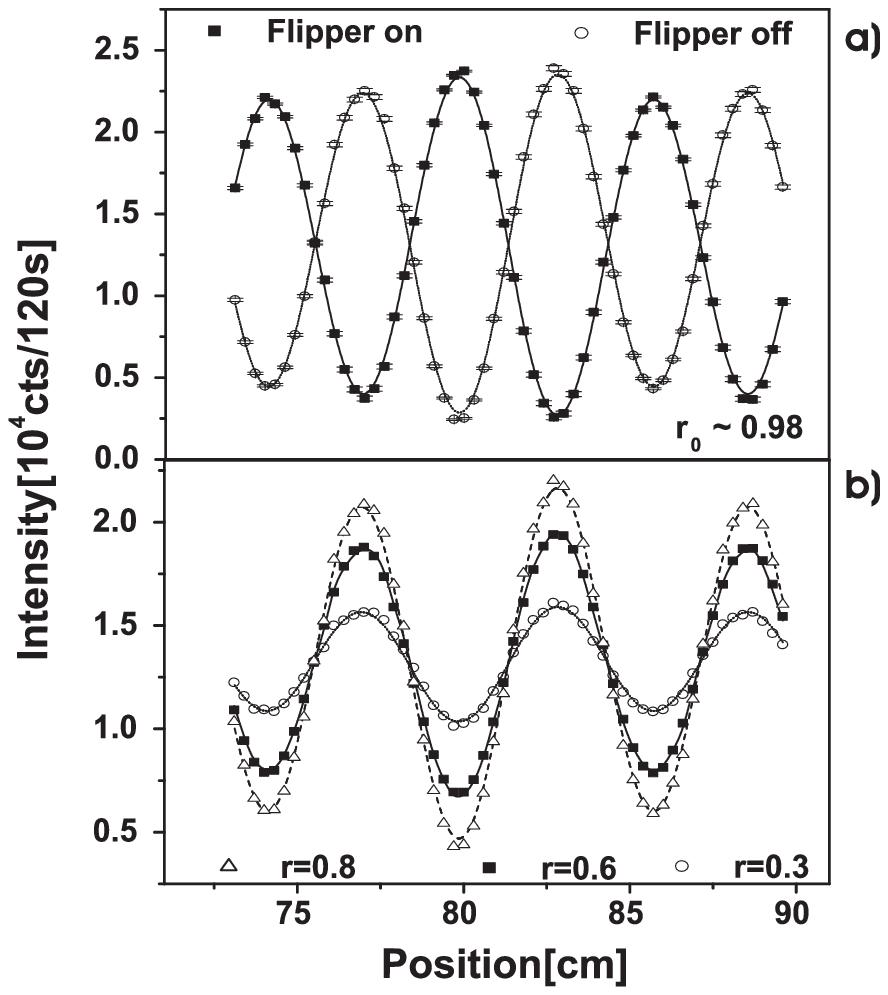}}
    \end{center}
    \caption{a) Measured intensities $\mbox{I}_{\mbox{\scriptsize off}}$ and
    $\mbox{I}_{\mbox{\scriptsize on}}$ for the parameter set
    $(\xi^{\mbox{\scriptsize (A)}},\delta^{\mbox{\scriptsize (A)}},
    \zeta^{\mbox{\scriptsize (A)}})\rightarrow(1.71,0.38,-1.46)$ rad at
    $r_0\sim0.98$.
    b) Calculated mixed state intensities for three degrees
    of polarization \emph{r}. Second order fits (see text) are included.}
    \label{fig3.eps}
\end{figure}
\begin{figure}
    \begin{center}
    \scalebox{0.7}{\includegraphics{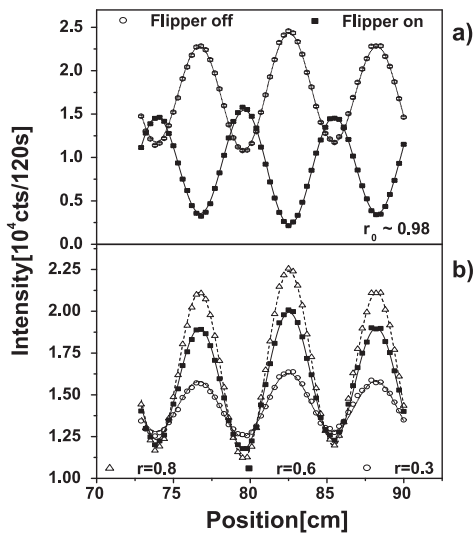}}
    \end{center}
    \caption{a) Measured intensities $\mbox{I}_{\mbox{\scriptsize off}}$ and
    $\mbox{I}_{\mbox{\scriptsize on}}$ for the parameter set
    $(\xi^{\mbox{\scriptsize (B)}},\delta^{\mbox{\scriptsize (B)}},
    \zeta^{\mbox{\scriptsize (B)}})
    \rightarrow (1.06,0.17,-1.40)$ rad
    at $r_0\sim 0.98$.
    b) Calculated mixed state intensities for three degrees of
    polarization \emph{r}. Second order fits (see text) are included.}
    \label{fig4.eps}
\end{figure}
The first example is a rotation as depicted schematically in Fig.
\ref{Fig2.eps}A. The associated SU(2) parameter set is
$(\xi^{\mbox{\scriptsize (A)}},\delta^{\mbox{\scriptsize
(A)}},\zeta^{\mbox{\scriptsize (A)}})\rightarrow(1.71,0.38,-1.46)$
rad. By application of Eq.(\ref{eq:MixedStatePhase}) one obtains
the theoretical result $\Phi_{\mbox{\scriptsize
th}}^{\mbox{\scriptsize (A)}}=0.37$ rad, whereas the value
computed from the measured data using
Eq.(\ref{eq:MixedStatePhInTermsOfImaxImin}) is
$\Phi_{\mbox{\scriptsize m}}^{\mbox{\scriptsize (A)}}=0.38\pm0.08$
rad at initial polarization $r^{\mbox{\scriptsize
(A)}}_0=0.976\pm0.004$. The measured intensities
$\mbox{I}_{\mbox{\scriptsize off}}$, $\mbox{I}_{\mbox{\scriptsize
on}}$ and the calculated mixed state intensities for three
different degrees of polarization are shown in Fig.
\ref{fig3.eps}a and b, respectively. The second example is
depicted in Fig. \ref{Fig2.eps}B, being described by parameters
$(\xi^{\mbox{\scriptsize (B)}},\delta^{\mbox{\scriptsize
(B)}},\zeta^{\mbox{\scriptsize (B)}})\rightarrow(1.06,0.17,-1.40)$
rad. For this set the theoretical prediction is
$\Phi_{\mbox{\scriptsize th}}^{\mbox{\scriptsize (B)}}=0.17$ rad.
The experimental result is $\Phi_{\mbox{\scriptsize
m}}^{\mbox{\scriptsize (B)}}=0.16\pm0.06$ rad at initial
polarization $r^{\mbox{\scriptsize (B)}}_0=0.981\pm0.005$.
Measured and calculated intensities are shown in Fig.
\ref{fig4.eps}a and b. For the density matrices derived from the
experimental data the results are $\Phi^{\mbox{\scriptsize
(A)}}_{0.8}=0.35 \pm 0.06~(0.31)$ rad, $\Phi^{\mbox{\scriptsize
(A)}}_{0.6}=0.21 \pm 0.06~(0.24)$ rad, $\Phi^{\mbox{\scriptsize
(A)}}_{0.3}=0.05 \pm 0.07~(0.12)$ rad for the transformation
$(\xi^{\mbox{\scriptsize (A)}},\delta^{\mbox{\scriptsize
(A)}},\zeta^{\mbox{\scriptsize (A)}})$ and
$\Phi^{\mbox{\scriptsize (B)}}_{0.8}=0.12 \pm 0.06~ (0.14)$ rad,
$\Phi^{\mbox{\scriptsize (B)}}_{0.6}=0.10 \pm 0.07~ (0.10)$ rad,
$\Phi^{\mbox{\scriptsize (B)}}_{0.3}=0.02 \pm 0.08~ (0.05)$ rad
for $(\xi^{\mbox{\scriptsize (B)}},\delta^{\mbox{\scriptsize
(B)}},\zeta^{\mbox{\scriptsize (B)}})$. The values in parenthesis
are the corresponding theoretical predictions for $\Phi$
calculated from Eq.(\ref{eq:MixedStatePhase}).
\section{\label{sec:disc}Discussion}
The extreme values of the intensity modulations show slight
irregularities in height as can be seen in the graphs. This can be
explained as an effect of second order neutrons (wavelength
$\lambda$/2). The monochromatization of the beam is governed by
the Bragg condition $2\mbox{d}\sin\theta_{\mbox{\scriptsize
B}}=\mbox{n}\lambda$, that - for a certain angle
$\theta_{\mbox{\scriptsize B}}$ - is fulfilled for n$=1,2,...$ .
Consequently, the beam used for the experiment is contaminated by
a percentage of neutrons with only half the intended wavelength.
In a time of flight (TOF) measurement, this percentage was found
to be $\lesssim$ 7.2 \% of the overall intensity. Neutrons that
travel at double velocity spend only half the time in magnetic
fields along the beam trajectory. Therefore, their spin rotates
through only half the originally intended angles, showing a
behavior rather different from the first order intensity. To
compute the relative phase, values for
$\mbox{I}_{\mbox{\scriptsize max}}$ and
$\mbox{I}_{\mbox{\scriptsize min}}$ were determined from a
least-squares fitting model (also shown in Fig. \ref{fig3.eps} and
Fig. \ref{fig4.eps}), taking into account the second order
neutrons. The careful reader will point out that randomness is
missing in the mixing procedure described above: the relative
frequencies for finding the system in either one of the pure
states $|+z\rangle$ or $|-z\rangle$ are not only well known, but
even determined by the experimenter. As a matter of fact, the
actual system, the neutron beam, is at no time of the experiment
in one of the mixed states associated with \emph{r}=0.8, 0.6 or
0.3. We are aware of the fact that the procedure carried out does
by no means constitute a general mixing method, but represents a
special approach within the wide scope of possible and accepted
techniques for mixed state preparation. Nevertheless, the density
matrices are generated by using the measured data resulting from
the conducted experiments, fulfilling the predictions developed by
Larsson and Sj\"{o}qvist \cite{LarssonSjoeqvist2003}. A neutron
optical experiment showing the \emph{r} dependence of the phase by
application of some other preparation method will be carried out
in the future.

The obtained experimental results are in good agreement with
theoretical predictions for the relative mixed state phase,
although systematic errors occur due to inherent difficulties of
the experiment. Even though the guide field was constructed
carefully, some inhomogeneity ($\sim0.4$\%) along its length
cannot be avoided completely, accumulating an error of
$\sim5^{\circ}$ in the angle $\eta$ along the distance 4 L$_0$.
Furthermore, it was difficult to compute values for the
readjustment of 4 L$_0$ after a change of SU(2) coil currents.
Consequently, the correct distance between the spin turners was
tuned by its variation at minimum intensity relative position of
spin turners and SU(2) coils, thereby exhibiting the maximum of
visibility. Another point is that the two $\pi/2$ spin turners
were situated within the guide field. Therefore they did not
exactly carry out the designated rotations of the polarization
vector to the -y direction and back to the +z direction. These
inaccuracies are intrinsic to this experiment and can hardly be
avoided at considerable effort. Further discussions gave rise to
interesting ideas for possible improvements of the experimental
setup concerning accuracy and robustness. At present, they are
tested and will be described in a forthcoming paper \cite{Sponar}.
\section{\label{sec:conc}Conclusion}
To summarize, we have measured the Pancharatnam relative phase for
the mixed state case in a neutron polarimetry experiment. The
dependence of the phase of the degree of polarization was
indicated by calculation of weighted sums of the density matrices
corresponding to the measured intensities. The experimental data
is in good agreement with theory. Deviations of the predicted
behavior were explained to arise due to contamination of the beam
by second order neutrons.

\end{document}